\documentclass[12pt,preprint]{aastex}
                                                                                
\shorttitle{Hard X-Rays and Optical Emission Lines in Local AGN}
\shortauthors{Heckman et al.}
                                                                                
\begin{document}
                                                                                

\title{The Relationship of Hard X-ray and Optical Line Emission
in Low Redshift Active Galactic Nuclei}

\author{T. M. Heckman\altaffilmark{1}, A. Ptak\altaffilmark{1},
A. Hornschemeier\altaffilmark{1,2},G. Kauffmann\altaffilmark{3}}

\email{heckman@pha.jhu.edu}

\altaffiltext{1}{Center for Astrophysical Sciences, 
Department of Physics \& Astronomy,
Johns Hopkins University,
Baltimore, MD 21218}
\altaffiltext{2}{Laboratory for High Energy Astrophysics, NASA Goddard Space 
Flight Center, Greenbelt, MD 20771}
\altaffiltext{3}{Max Planck Institut fur Astrophysik, D-85748 Garching, Germany}

 

\begin{abstract}
In this paper we assess the relationship of the population of
Active Galactic Nuclei (AGN) selected by hard X-rays to the 
traditional population of AGN with strong optical emission lines.
First, we study
the emission-line properties of a new hard X-ray
selected sample of 47 local AGN (classified optically as both Type 1 and Type 2
AGN). We find that the hard X-
ray (3-20 keV) and [OIII]$\lambda$5007 optical emission-line
luminosities are well-correlated over a range of about four
orders-of-magnitude in luminosity (mean luminosity ratio
2.15 dex with a standard deviation of  $\sigma$ = 0.51 dex).
Second, we study the hard X-ray properties of a sample of 55
local AGN selected from the literature on the basis of the flux in
the [OIII] line. The correlation between the hard X-ray (2-10 keV) and
[OIII] luminosity for the Type 1 AGN is consistent with
what is seen in the hard X-ray selected sample. However, the
Type 2 AGN have a much larger range in the luminosity
ratio, and many are very weak in hard X-rays (as expected
for heavily absorbed AGN). We then compare the hard X-ray (3-20
keV) and [OIII] luminosity functions of AGN in the local
universe. These have similar faint-end slopes with a
luminosity ratio of 1.60 dex (0.55 dex smaller than the mean
value for individual hard X-ray selected AGN). We conclude that 
at low redshift, selection by narrow optical emission-
lines will recover most AGN selected by hard X-rays (with the
exception of BL Lac objects). However,
selection by hard X-rays misses a significant fraction
of the local AGN population with strong emission lines.
\end{abstract}
 
 

 

\section{Introduction}

One of the fundamental properties of Active Galactic Nuclei
(AGN) that sets them apart from stars and galaxies is the
very broad-band nature of their intrinsic spectral energy
distribution. In Type 1 AGN where we
have a clear view of the ``central engine'', there is a
comparable amount of energy radiated per decade in frequency
over a span of more than six orders-of-magnitude in photon
energy from the far-infrared to hard X-rays (e.g. Elvis et
al. 1994).  The panchromatic nature of AGN is both a
blessing and a curse. It enables us to find and then
investigate AGN with a broad array of facilities, but this
very diversity implies that any single
technique may provide a biased perspective (e.g. Mushotzky 2004).
In this paper
we address this issue in the context of local samples of AGN
selected by their hard X-ray emission, compared to AGN
with strong optical emission-lines.

This particular comparison is very timely.
On the one hand, ``mega surveys'' like the 2dF Galaxy Redshift
Survey (Colless et al. 2001),  the 2dF QSO Redshift Survey
(Croom et al. 2004), and the Sloan Digital Sky Survey (SDSS
-York et  al. 2000) are generating samples of hundreds of
thousands of AGN identified on the basis of their optical
spectroscopic properties. These data not only allow AGN to
be efficiently recognized (e.g. Kauffmann et al 2003; Hao et
al. 2005a; Zakamska et al. 2003), 
but the same data can be used to quantify many
of the key properties of the AGN and their host galaxies
(e.g. Kauffmann et al. 2003). Recently,  Heckman et al.
(2004) have used the SDSS data to show that the population
of low mass black holes in the present day universe is
growing at a substantial rate, and that the co-construction
of black holes and bulges is linked today. To what extent
are these results affected by the use of optical emission
lines to identify and quantify the AGN?  For example, radio-
selected AGN in the SDSS are associated with a very
different population (the most massive black holes and giant
elliptical galaxies - Best et al. 2005a,b).

On the other hand, the deep X-ray surveys
carried out by the Chandra X-ray Observatory have
revolutionized our picture of the cosmic history of the AGN
phenomenon and thus of the growth of supermassive black
holes (e.g. Ueda et al. 2003; Barger et
al. 2005 - hereafter B05; Hasinger et al. 2005).
These results show that a
substantial fraction of the growth of supermassive black
holes as traced by hard X-ray emission has occurred
relatively recently (since $z \sim$ 1), and that this late
growth has preferentially occurred in AGN of low or
moderate luminosity. It is reassuring that these results
appear compatible with the inferences drawn from the SDSS
sample of emission-line AGN, at least qualitatively.
However, the relationship between AGN selected via hard X-
ray emission and optical emission lines is still murky.

While the most luminous hard-X-ray-selected AGN are usually
identified optically with quasars, most of the lower luminosity
objects are not. There have been several explanations for
this. The most revolutionary possibility (Steffen et al.
2003; B05) is that these lower luminosity objects
are a genuinely new type of AGN
that intrinsically lacks the strong UV and optical continuum
and associated Broad Line Region that defines Type 1 AGN.
The second and most mundane explanation is that this is just
``aperture dilution'': at the redshifts of these AGN the slit
of the optical spectrograph encompasses a large fraction of the
entire AGN host galaxy and the observed light is dominated
by the host galaxy except in the case of the most powerful
AGN. This makes it difficult to recognize the presence of an
AGN (Moran, Filippenko, \& Chornock 2002).

The final possibility (e.g. B05) is that these are normal AGN that
are dust-obscured at optical and ultraviolet
wavelengths, but that this obscuring material is easily
penetrated by hard X-rays (absorbing column $< 10^{23}$ cm$^{-2}$).
The widespread
existence of obscured AGN has been known since it was
realized over twenty years ago (Lawrence \& Elvis 1982;
Antonucci \& Miller 1985; Krolik
\& Begelman 1988) that Type 1 and Type 2 AGN
could be unified.
In a Type 1 (Type 2) AGN our line of sight is near the
polar axis (equatorial plane) of a dusty toroidal structure
that surrounds the  central engine (black hole plus
accretion disk). The torus is optically-thick to visible and
ultraviolet photons emitted by the central engine, but (in
many AGN) is optically thin to hard X-rays (e.g. Risaliti, Maiolino,
\& Salvati 1999). To reconcile this picture with the results from
the hard X-ray surveys would require that the ``covering
fraction'' of the obscuring torus increases systematically with
decreasing hard X-ray luminosity (Ueda et al. 2003; B05).

It has been difficult to decisively discriminate between the
above possibilities, largely because the samples of hard X-
ray selected AGN have mostly come from deep surveys that
cover relatively small solid angles. As a result the AGN and
their host galaxies are relatively faint and distant, making
it difficult to investigate them in detail.

In this paper we will use existing catalogs to compare the
properties of AGN selected by hard X-rays to those with bright optical
emission lines. In section 2 we will describe a new ``all-sky''
sample of 95 relatively nearby AGN detected serendipitously
in the 3-20 keV hard X-ray band in the Rossi X-ray Timing
Experiment slew survey (Revnivtsev et al. 2004; Sazonov \&
Revnivtsev 2004) and also a sample derived from 
extensive
compilations of AGN [OIII]$\lambda$5007 emission-line properties (Xu,
Livio, \& Baum 1999; Whittle 1992).
In section 3 we will compare the relationship
between the hard X-ray and [OIII] luminosities for these two samples.
In section 4 we will compare the hard X-
ray and [OIII] luminosity functions for AGN in
the local universe. In section 5, we will examine the
dependence of the ratio of Type 1 and Type 2 AGN on their
hard X-ray luminosity.  Finally, we will summarize our
results and their implications in section 6.

\section{AGN Samples}

\subsection{AGN Selected by Hard X-rays}

Sazonov \& Revnivtsev (2004 - hereafter SR04) have provided
optical identifications and hard X-ray luminosities (in the
3 to 20 keV band) for 95 AGN detected in the RXTE slew
survey (XSS). The XSS covered 90\% of the sky at $|b| >
10^{\circ}$ to a flux limit in the 3-20 keV band of  $2.5 \times 10^{-11}$ erg
s$^{-1}$ cm$^{-2}$ or better. There are 35 sources in the
XSS that are not identified optically, and based on their X-
ray spectra these are likely to be AGN.  The XSS provides
positions to only $\sim 1^{\circ}$ accuracy, so additional
follow-up hard X-ray observations will be required to
identify these sources.  In the Northern celestial
hemisphere where the optical catalogs used for cross-
identification are more extensive, there are 45 identified
AGN and only 4 unidentified sources. Thus, the SR04 sample
provides a reliable census of the sources that dominate the
AGN population at relatively bright hard X-ray flux
levels.

The SR04 sample has a median redshift of $z \sim$ 0.035, but extends out
to $z$ = 2.70. Since our
goal in this paper is to compare AGN populations in the local universe,
we restrict the SR04 sample
to $z < 0.2$. This
eliminates only eight objects. The sample then
consists of 65\% Type 1 AGN (Type 1 Seyfert nuclei, quasars,
and broad line radio galaxies), 22\%, Type 2 AGN (Type 2
Seyferts and narrow line radio galaxies), and 13\% BL Lacs.
Given that 92\% of the northern sample is identified, one
immediate conclusion is that at most 8\% of the hard X-ray
selected objects would not be otherwise recognized as
AGN. We emphasize that because of the low median redsift of this sample,
aperture effects in the optical spectra will be small (e.g. Kewley,
Jansen, \& Geller 2005; Kauffmann et al. 2003).
The optical spectroscopic signature of the AGN
would be more difficult to detect if these AGN were at higher redshift.

\subsection{AGN with Bright [OIII] Emission Lines}

In the standard ``unified model'' for AGN (e.g.
Antonucci 1993; Urry \& Padovani 1995),
narrow (FWHM $\sim$ several hundred km/s) optical emission lines arise in gas 
located
several hundred pc or more from the central engine and are
excited by ionizing radiation escaping along the polar axis
of the obscuring torus. Since this material lies outside the
region of high gas column density (the torus), the narrow optical emission
lines suffer only moderate amounts of dust obscuration
(e.g. Dahari \& De Robertis 1988; Keel et al. 1994; Kauffmann et al. 2003), 
even in cases in which the central hard X-ray source is almost totally obscured
by a Compton-thick torus (e.g. Bassani et al. 1999).
These narrow emission lines should therefore provide an unbiased
orientation-independent indicator of the ionizing luminosity
of the central engine in both Type 1 and Type 2 AGN.

This concept has been validated empirically in several
different studies of AGN (e.g. Mulchaey et al.
1994; Heckman 1995; Keel et al. 1994).  As discussed
extensively in Kauffmann et al. (2003), Heckman et al. (2004), and
Brinchmann et al. (2004), the
[OIII]$\lambda$5007 line is the best optical estimator of the AGN
power in Type 2 AGN, as it is the strongest narrow emission
line in such AGN and the least contaminated by contribution
from HII regions associated with star formation in the host
galaxy.

Unfortunately, there is no all-sky catalog of AGN selected purely on the
basis of their [OIII]$\lambda$5007 flux (analogous to the SR04 hard X-
ray sample). The best that we can do is to select from the known
population of local AGN those objects with the brightest [OIII]
emission lines. 
Xu, Livio, \& Baum (1999 - hereafter X99) have published 
a catalog of [OIII]$\lambda$5007 luminosities for 409 AGN. We have supplemented
this with the similar catalog of 140 AGN compiled by Whittle (1992
- hereafter W92). These catalogs are not complete, and their members
were originally discovered on the basis of a wide range of properties.
While they should be representative of the [OIII]-bright local AGN 
population, there may well be biases in this sample compared
to a ideal complete sample selected purely on the basis of [OIII] flux.

\section{Comparison of Hard X-ray and [OIII] Properties}

\subsection{AGN Selected by Hard X-ray Emission}

Of the 87 hard X-ray selected AGN in the SR04
catalog with $z < 0.2$, the union of the W92 and X99
optical emission line catalogs contains
47 (54\%).
These are listed in Table 1. The fraction of the SR04 sample in the
merged X99 plus W92 catalogs is much higher in the northern sky (28 of 40, or
70\%) compared to the south (19 of 47, or 40\%).

While the SR04 Type 1 and Type 2 AGN populations are well represented in
Table 1,
this is not the case with the BL Lacs.
BL Lacs comprise 13\% of the SR04 sample, but none of these have [OIII]
data in either X99 or W92. From the definition of a BL Lac object (e.g.
Wolfe 1978),
these will have generally have much weaker optical line emission than 
any of the other AGN classes in the SR04 catalog. We will keep this
in mind in the discussion that follows. However, if
we specifically exclude the BL Lac objects,
then the fraction of AGN in the SR04 northern sample with [OIII]
data in X99 or W92 is very high (28 of 33, or 85\%). 
The corresponding fraction in the
south is 19 or 43, or 44\%. 

Therefore, the high degree
of completeness in the SR04 northern sample (92\% identified - see above)
combined with the large fraction of these identified sources with [OIII]
fluxes (85\%)
makes us confident that the northern component of the sample listed in 
Table 1
is representative of the optical properties of the local hard X-ray
selected AGN population (with the important exception of BL Lac objects).

In Figure 1 we plot a
histogram of the ratio log $(L_{HX}/L_{OIII})$ for the AGN in Table 1. The
distribution has a mean of 2.15$\pm$0.08 dex (a factor of
140) and a standard deviation of 0.51 dex (a factor of 3.3).
In Figure 2 we plot the hard X-ray {\it vs.} the [OIII]$\lambda$5007
luminosity for this sample. \footnote{We have converted all luminosities to
a cosmology with $H_0$ = 70 km s$^{-1}$ Mpc$^{-1}$.} 
Note that in the cases of both [OIII] and hard X-rays, we have used
the observed fluxes rather than fluxes corrected for intrinsic absorption.
This is motivated by the main intent of our paper, which is to
compare AGN selected on the basis of their 
{\it observed} properties. This is discussed further in the Appendix below.

By way of comparison, the distribution of the
ratio of the optical continuum and [OIII] luminosity for
optically selected Type 1 AGN in the
SDSS has a standard deviation of 0.34 dex (see 
Heckman et al. 2004; Zakamska et al. 2003), while
the distributions of the ratio of the 60 $\mu m$ and [OIII]
luminosities for far-IR-selected Type 1 and Type 2 AGN
both have standard deviations of $\sim$0.5 dex (Keel et al.
1994). 

To test the robustness of the result shown in Figures 1 and
2 we have subdivided the SR04 sample into Type 1 and Type 2
AGN. We find no significant difference in mean of the
luminosity ratio (2.14 dex for Type 1 AGN {\it vs.} 2.17 dex
for Type 2 AGN).  As emphasized above, the SR04 sample is
significantly more complete in the northern sky than in the
south. Subdividing the sample into celestial hemispheres, we
again find no significant difference in the mean luminosity
ratio (2.11 dex in the north {\it  vs.} 2.17 dex in the
south).

\subsection{AGN with Bright [OIII] Emission Lines}

To construct a local AGN sample with a depth
similar to the SR04 hard X-ray sample, we selected all
AGN in the X99 or W92 catalogs with $z <$ 0.2 and an [OIII]$\lambda$5007 flux
greater
than $2.5 \times 10^{-13}$ erg cm$^{-2}$ s$^{-1}$ (100 times
lower than the limiting hard X-ray flux in SR04).
All the AGN selected lie in the SR04 footprint on the sky ($|b| >
10^{\circ}$). This results in a sample of 55 AGN
listed in Table 2 (hereafter the [OIII]-bright sample).

If the hard X-ray selected sample discussed in
section 3.1 were representative of the entire AGN population,
we would expect a majority of the AGN in Table 2 to be detected
in hard X-rays at the flux limit of the SR04 catalog. More quantitatively,
folding the [OIII] fluxes for the sample in Table 2 through the
HX/[OIII] flux ratio distribution for the hard X-ray selected
SR04 sample in Figure 1 predicts that $\sim$70\% of the [OIII]-bright
sample (38 objects) should be detected in hard X-rays at the limit of the SR04
catalog. Instead, only 42\% (23 objects) are
detected. The lower-than-expected yield is entirely due to
the Type 2 AGN (7 of 32, or 22\%),  while the Type 1 AGN
are detected at the expected rate (16 of 23, or 70\%). 
The missing Type 2 AGN are almost certainly heavily absorbed
($N_H \sim 10^{23}$ to $10^{24}$ cm$^{-2}$) or even Compton-thick 
($N_H > 10^{24}$ cm$^{-2}$). This is consistent with the statement by
SR04 that their sample contains no Compton-thick AGN,
which Bassani et al. (1999) show comprise a significant fraction
of Type 2 AGN.

To further investigate the hard X-ray properties of the
sample, we have used literature compilations of X-ray fluxes
in the 2-10 keV band (X99, Bassani et al. 1999),
supplemented when necessary by our own analysis of archival
data. This allows us to reach significantly lower hard X-ray
fluxes than the SR04 sample, albeit in a somewhat lower energy
band. These data are available for 49
of the 55 [OIII]-selected AGN (6 have no published or
archival hard X-ray data). 

The effects of photoelectric absorption are potentially more
significant in the 2-10 keV band than in the 3-20 keV RXTE band
in the SR04 sample. As stated above, our goal in this paper is
to compare samples of AGN selected on the basis of their observed
properties (e.g. hard X-ray or [OIII] flux). Thus, neither the [OIII] nor
the 2-10 keV fluxes we use have been
corrected for absorption. 
For Type 2 AGN (which can often
have substantial absorbing column densities) we know that we are
underestimating the intrinsic hard X-ray flux. This underestimate can
be as large as a factor of $\sim10^2$ for Compton-thick AGN 
(e.g. Comastri 2004). 
We address this issue in more detail in the Appendix.

The results are shown in Figures 3 and 4. The relationship
between hard X-ray and [OIII] luminosities is well-behaved
for the Type 1 AGN (mean ratio 1.59 dex, $\sigma$ = 0.48
dex). Based on the 23 AGN in Table 2 with {\it both} 3-20 keV and 2-10
keV fluxes, the corresponding luminosity ratio would be
1.96 dex in the 3-20 keV band (similar to the value 2.14 dex
for Type 1 AGN in the SR04/X99 sample above). A minority of
the Type 2 AGN are bright in hard X-rays and overlap the
range in the luminosity ratio defined by the Type 1 AGN.
However, the Type 2 AGN as-a-whole are significantly weaker
in hard X-rays, and the range in the HX/[OIII] luminosity
ratio is much larger (mean = 0.57
dex and $\sigma$  = 1.06 dex) compared to the Type 1 AGN. 

We therefore
conclude that while the hard X-ray {\it vs.} [OIII] properties of Type 1 AGN are
relatively insensitive to the selection criteria, most optically
selected Type 2 AGN are significantly weaker in hard X-rays than their
counterparts in a hard X-ray selected sample. This is qualitatively
consistent with earlier investigations of Type 2 AGN,
which documented the existence of a significant
population of X-ray-faint AGN
(e.g. Bassani et al. 1999;
Risaliti, Maiolino, \& Salvati 1999). We emphasize again that this does
not imply that these AGN are {\it intrinsically} weak
hard X-ray emitters. Instead, they are heavily absorbed
or even Compton-thick AGN in which only
a small fraction of the emitted hard X-rays are observed
(Comastri 2004; Levenson et al. 2002). 

\section{Comparison of Hard X-ray and [OIII] Luminosity
Functions}

We have shown above that there is a good correlation between
the [OIII] and hard X-ray luminosities for a hard X-ray
selected sample of AGN.  However, this is not the case for
sample of AGN with bright [OIII] line emission: many of the Type 2 AGN
in this sample are
far weaker in hard X-rays than Type 1 AGN with the same
[OIII] luminosity. To gain insight into which of the samples
is more representative of the local AGN population, in the
present section we compare the independently determined hard
X-ray and [OIII] luminosities functions for AGN in the local
universe.

SR04 have used the XSS to derive a local ($<z>$ = 0.034)
hard X-ray (3-20 keV) luminosity function for Compton-thin
AGN (excluding BL Lac objects). Hao et al. (2005b) used a local ($<z>
\sim$ 0.1) AGN sample extracted from the SDSS to derive the
[OIII]$\lambda$5007 luminosity function. We compare these two
luminosity functions in Figure 5.
\footnote{We have
used the evolutionary parameters measured by B05 to slightly adjust
the SR04 luminosity function to the mean redshift of the
SDSS sample.} 

The hard X-ray and [OIII] luminosity functions have similar
slopes. To align the two functions in luminosity at fixed
space density requires an offset in the luminosity scale (HX/OIII) by
1.53 to 1.68 dex (depending the assumed incompleteness
correction applied to the SR04 sample - see SR04 for details). 
This ratio is significantly less than the mean
value we have measured above for log $(L_{HX}/L_{OIII})$ =
2.15 for individual hard X-ray selected AGN in the SR04 sample
itself. Alternatively, if we fix the luminosity ratio at
2.15 dex, then at the same equivalent luminosity the space density
of hard X-ray selected AGN is lower than that of [OIII]-selected AGN
by a factor of $\sim$0.5 dex.
These results are consistent with a significant population of
[OIII]-bright AGN that are relatively weak in the hard X-ray band.

\section{Luminosity Dependent Properties}

Deep X-ray surveys find that
objects classified optically as Type 1 AGN are a minor
component of the hard
X-ray selected AGN population at low luminosities (below
$L_{HX} \sim10^{43.5}$ erg/s). As noted in the introduction, there
have been several different interpretations of this result, ranging
from the emergence of a new type of AGN to the effects of aperture dilution
in the optical spectra.
Since the AGN in the SR04 sample are so much nearer and brighter,
they provide an opportunity to test these interpretations.

Notably, the demographics of the SR04 sample are very different than
in the B05 deep X-ray survey: 
for $L_{HX} < 10^{43.5}$ erg/s, the SR04 sample
contains roughly equal numbers of optically-classified
Type 1 and Type 2 AGN 
(18 {\it vs.} 17 objects). Could this difference be the result
of cosmic evolution in the AGN population? If so, this evolution
must have occurred very recently, since
the apparent lack of low luminosity Type 1 AGN
in the deep surveys persists even in the
lowest redshift range investigated in B05 ($z$ = 0.1 to 0.4).
\footnote{We emphasize that the B05 and SR04 samples do agree with one another 
at higher X-ray luminosities, where both are dominated by AGN
classified optically as Type 1.}

The difference in the Type 1 AGN fraction between SR04 and the deep 
X-ray surveys is
consistent with the idea that many low
luminosity Type 1 AGN are not recognized as such in the deep
surveys because of severe dilution of the optical
spectra by light from the host galaxy 
(Moran, Filippenko, \& Chornock 2002; Severgnini et al. 2003).
Moreover,
Silverman et al. (2005) also show that whether or not a hard X-ray source
is classified optically as a Type 1 AGN can depend sensitively on which
portion of the rest-frame spectrum is observed (e.g. a Broad Line Region
is much easier to recognize in the rest-frame red (with H$\alpha$) or 
near-UV (with MgII$\lambda$2800) than at intermediate wavelengths
where H$\beta$ is the strongest line).

B05 argue against the aperture-dilution explanation by using
the high spatial resolution of HST imaging to measure a
``nuclear'' flux at 3000\AA\ (rest frame). They show that the
ratio of nuclear UV to hard X-ray flux is significantly
higher {\it in the mean} in objects classified optically as
Type 1 AGNs than in the others. However, there is a very
wide range in $F_{UV}/F_{HX}$ among the objects 
classified optically as Type 2 AGNs ($\sim$ 3 dex), and this range
overlaps that of the Type 1 AGNs. It is possible
that the subset of the Type 2 AGN with high values of
$F_{UV}/F_{HX}$ are in fact misclassified Type 1 AGN.

\section{Summary}

In this paper we have explored the relation between hard X-
ray emission and the [OIII]$\lambda$5007 emission line in local
samples of AGN. We have used the SR04 hard X-ray (3-20 kev)
selected sample of AGN, the X99 and W92 catalogs of [OIII]$\lambda$5007 AGN
luminosities, and the Hao et al. (2005b) and S04 determinations of AGN
emission line and hard X-ray luminosity functions respectively to draw
the following conclusions:

\begin{itemize}

\item
For the SR04 hard X-ray selected sample of Type 1 and Type 2
AGN, the hard X-ray luminosity is well correlated with the
[OIII]$\lambda$5007 luminosity over a range of $\sim$ four orders-of-
magnitude in luminosity. The mean value for
log $(L_{HX}/L_{OIII})$ is 2.15 dex ($\sigma$ = 0.51 dex).
The Type 1 and Type 2 AGN  have the same mean HX to [OIII]
luminosity ratio (as expected in the unified model for AGN).
BL Lac objects - which comprise 13\% of the SR04
sample - are not included in this comparison. They would have much weaker [OIII]
emission.

\item
For a sample of [OIII]-bright AGN derived from the X99 and W92
catalogs, the Type 1 AGN follow a hard X-
ray {\it vs.} [OIII] luminosity relation similar to the hard X-ray
selected AGN sample. However, most of the [OIII]-bright
Type 2 AGN  are weaker hard X-ray emitters (by an average of
1.02 dex in the 2-10 keV band) and the scatter in the
HX/[OIII] luminosity ratio (1.06 dex) is much larger than for the Type
1 AGN (0.48 dex).

\item
The [OIII]$\lambda$5007 luminosity function derived for optically-
selected AGN from the SDSS agrees with the hard X-ray
luminosity function derived for hard X-ray selected AGN for
an HX/[OIII] luminosity ratio at a given space density of
1.60 dex. This is significantly smaller than the mean
HX/[OIII] luminosity ratio of 2.15 dex for AGN in the SR04 sample.
Alternatively - for a luminosity ratio of 2.15 dex - the
space density of hard X-ray selected AGN is $\sim$0.5 dex smaller
than [OIII] selected AGN of the same equivalent luminosity.

\item
At low hard X-ray luminosities ($<10^{43.5}$ erg/sec),
the relative numbers of optically-classified Type
1 and Type 2 AGN in the SR04 hard X-ray selected sample are
similar (while Type 1 AGN dominate at higher luminosities).
The existence of a substantial population of
Type 1 AGN with such low X-ray luminosities is not
consistent with results from deep X-ray surveys. Either there has
been strong cosmic evolution since $z \sim$ 0.1 - 0.4, or many
low luminosity Type 1 AGN are being misclassified in the deep
surveys due to dilution by galaxy light in the optical spectra.

\end{itemize}

As noted in the introduction, the use of both hard X-
ray and [OIII] data allows inferences to be made about the
cosmic evolution of the AGN population (and its relationship
to black hole growth and galaxy evolution) that in principle are more
robust and physically instructive than either type of data
provide in isolation.

In this context, the results above can be viewed as the classic case
of a glass half full or a glass half empty.
They do imply that (at least in the low-
redshift universe) samples of AGN with bright [OIII]$\lambda$5007
emission are not missing a significant fraction of the
AGN selected by hard X-ray emission. Specifically, while
BL Lac objects {\it are} weak [OIII] emitters, they comprise a 
distinct minority population of X-ray bright AGN
(13\% in SR04).
The ``X-ray bright,
optically normal galaxies''  (Levenson et al. 2001;
Comastri et al. 2002; Maiolino
et al. 2003; Brusa et al. 2003) are evidently not a major component of
the local AGN population. This is consistent with Silverman et al. (2005)
who find they comprise only 7\% of the CHAMP sample (and see also
Hornschemeier et al. 2005).

However (at least in the low-
redshift universe), selection by hard X-
rays does miss a significant fraction of the AGN population
selected by the [OIII]$\lambda$5007 optical emission-line. This
population of Type 2 AGN has been long recognized (e.g.
Risaliti, Maiolino, \& Salvati 1999; Bassani et al. 1999), and interpreted as
cases in which the hard X-ray source is heavily obscured and even 
Compton-thick
(e.g. Levenson et al. 2002; Comastri 2004). 

It is certainly possible
that the class of heavily-absorbed AGN might evolve differently with
cosmic time and/or trace a different population of black holes
and their host galaxies and/or represent a distinct phase in the
evolution of individual AGN (e.g. Wada \& Norman 2002; Stevens et al. 2005).
A significant population of Compton-thick AGN is required to match
the overall spectrum of the cosmic X-ray background at energies above
$\sim$10 keV (e.g. Comastri 2004). Thus, despite
the obvious advantages of hard X-ray surveys (e.g. Mushotzky 2004),
it seems that multiwaveband investigations are necessary
to provide a complete picture of the cosmic evolution of
the AGN population and its relationship to galaxy evolution.
 
 
\acknowledgments

We thank an anonymous referee for helping us to strengthen this paper.
We acknowledge the support of NASA grants NNG04GE47G and NNG04GF79G.

\appendix
\section{X-ray Fluxes for Absorbed AGN}

Given the relatively low energy resolution of typical detectors used
in X-ray astronomy, fluxes
are derived in a model dependent way. A parameterized
model for the X-ray spectrum is folded through the
known response of the detector and best-fit model parameters are
determined through comparison to the data (e.g. Arnaud 1996).
In the case of AGN, the 
model most commonly used is an intrinsic power law spectrum transmitted 
through a column density of cold gas with solar relative elemental abundances.
The model fit then yields the power-law index, the absorbing column,
and the flux. Fluxes are sometimes reported as-observed,
and these are the fluxes listed in Table 2 and used to make Figures
3 and 4 (e.g. Bassani et al. 1999). Fluxes are also  
commonly reported after applying the correction for
foreground absorption derived from the model fit 
(sometimes called ``unabsorbed fluxes''). 

The procedure to derive the unabsorbed fluxes are adequate for Type 1 AGN in
the hard X-ray
band since the absorbing columns and resulting reductions in transmitted 
flux are generally very small (see for example SR04).
However, the methodology is often inadequate in the
cases of those Type 2 AGN where absorbing columns often exceed 
$10^{23}$ cm$^{-2}$, and can even be Compton-thick ($>10^{24}$ cm$^{-2}$).
In these cases, the emerging spectrum will show the
complex effects of reprocessing and radiative transfer. In well-studied
cases with good X-ray spectra (e.g. Turner et al. 1997; Ogle et al. 2003),
successful models require absorption
by an inhomogeneous absorber (``partial covering models''), and/or reprocessing
of X-rays illuminating cold or partially ionized gas, and/or electron
scattering of X-rays escaping along the polar axis of the torus. 
When the X-ray emission from the AGN is heavily absorbed, emission from
the host galaxy is often significant and adds further spectral complexity
(e.g. Levenson et al. 2005). 
Unabsorbed fluxes derived from fitting
such complex spectra with a simple absorbed power law model can
significantly underestimate the true intrinsic flux - by up to a factor
of $\sim10^2$ for Compton-thick AGN (e.g. Comastri 2004).

To illustrate this, we have retrieved archival XMM-Newton X-ray spectra
for 12 of the Type 2 AGN in Table 2 (NGC 526A, NGC 1068, NGC 1386, NGC 2992,
NGC 4388, NGC 4507, NGC 5506, NGC 7212, NGC 7582, Mrk 1, Mrk 3, and
Tol 0109-383). We have fit each spectrum with two types of models. The
first is a simple absorbed power law. The second is a partial covering model
in which we fit for both the absorbing column and covering fraction
of the absorber. We have also added a narrow Fe K$\alpha$ line to the
model, as this line can have a very large equivalent width in Compton-thick
AGN (e.g. Levenson et al. 2002). This is meant to be a more
physically realistic model for heavily absorbed AGN. For these AGN we
find that the mean absorption correction to the hard X-ray luminosity
derived from the simple power law fits is only 0.08 dex. In contrast,
the mean absorption correction derived from the partial covering models
is substantially larger (0.51 dex). Moreover, the
intrinsic 2-10 keV luminosity 
as derived from the partial covering model is larger than that
measured from the simple powerlaw model
by 0.61 dex on-average. 

The main conclusion then is that much of the
large scatter seen in the ratio of the observed hard X-ray 
to [OIII] fluxes in the Type 2 AGN in the [OIII]-bright sample 
(Figures 3 and 4)
is almost certainly an artefact of absorption of the hard X-rays
(see also Bassani et al. 1999). 
A related conclusion is that fitting simple absorbed power law models
to heavily absorbed AGN significantly underestimates the intrinsic
hard X-ray luminosity.

However, we emphasize that this does not
affect the point we are making in this paper: we are 
interested in empirically
assessing the effect of the bias of hard X-ray selected samples 
against AGN with weak {\it observed} hard X-ray emission. From this
perspective it does not matter whether the weak X-ray emission
is an intrinsic property of the AGN or just the result of severe absorption
of hard X-rays. If severe enough, then it can not be properly corrected
using the
simple models appropriate to fitting the typical ``survey-quality'' 
X-ray spectra of AGN.

 

 
\begin{figure}
\epsscale{.80}
\plotone{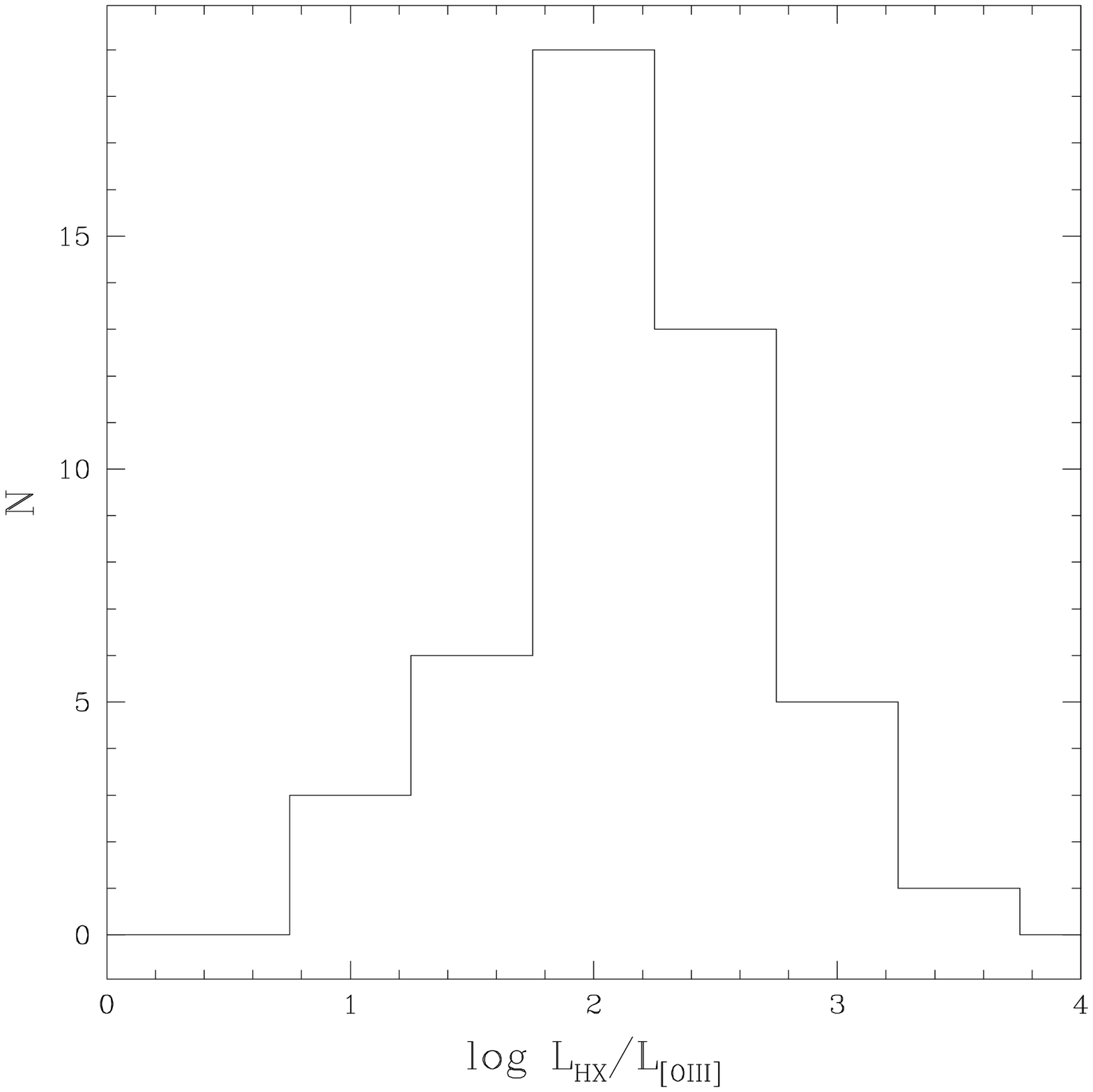}
\caption{A histogram of the log of the ratio of the hard X-ray (3-20 keV)
to [OIII]$\lambda$5007 luminosities for a sample of 47 local AGN selected
on the basis of their hard X -ray flux (the SR04 sample). The distribution
has a mean of 2.15 dex and a standard deviation of 0.51 dex. There is
no significant difference between the Type 1 and Type 2 AGN in this sample
(see text for details).}
\end{figure}

\begin{figure}
\epsscale{.80}
\plotone{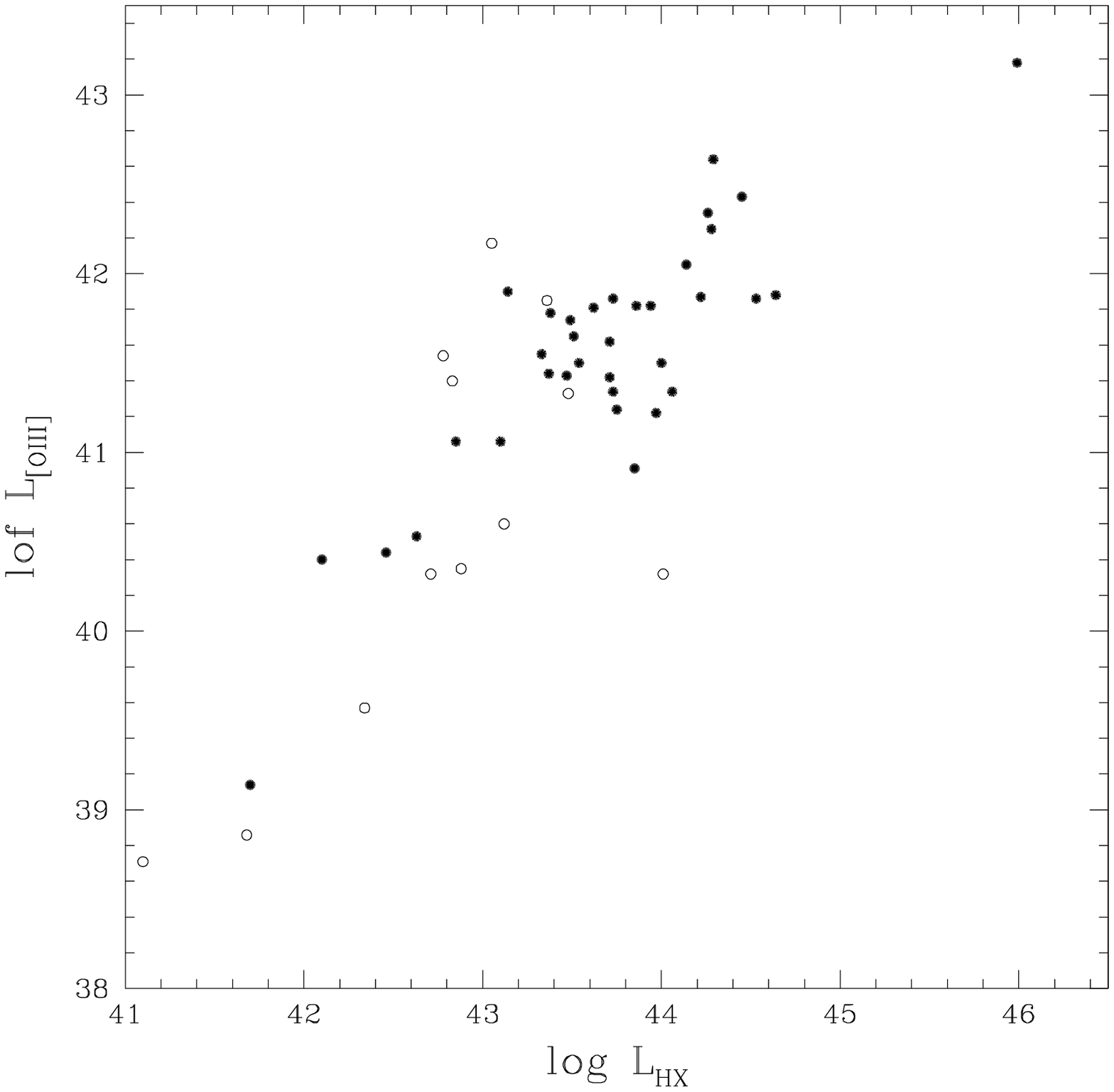}
\caption{A plot of the hard X-ray (3-20 keV) {\it vs.} the [OIII]$\lambda$5007
luminosities for the AGN in Figure 1. The Type 1 AGN are plotted as filled
circles and the Type 2 AGN as hollow circles.
Luminosities are in units of erg/sec.}
\end{figure}

\begin{figure}
\epsscale{.80}
\plotone{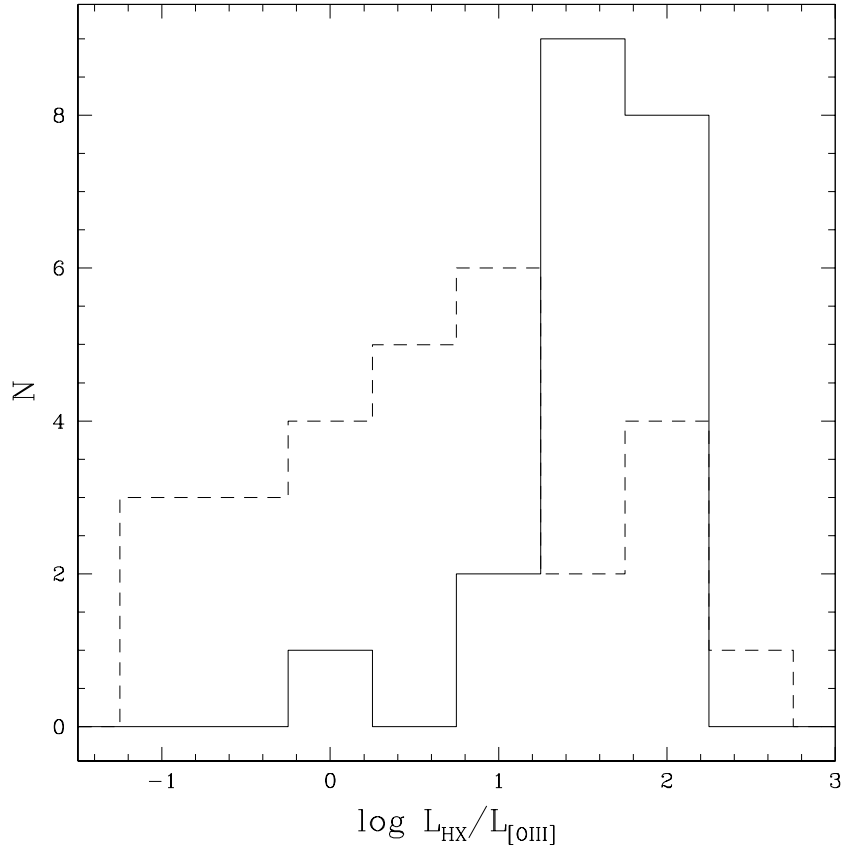}
\caption{Histograms of the log of the ratio of the hard X-ray (2-10 keV)
to [OIII]$\lambda$5007 luminosities for a sample of 49 local AGN selected
on the basis of their [OIII] flux (from the X99 and W92 catalogs). There is a 
major
difference between the Type 1 AGN (solid line) and Type 2 AGN (dashed line) 
in this sample. For the Type 1, the mean is 1.59 dex with a standard deviation
of only 0.48 dex, while for the Type 2 AGN the mean is only 0.57 dex with
a standard deviation of 1.06 dex.}
\end{figure}

\begin{figure}
\epsscale{.80}
\plotone{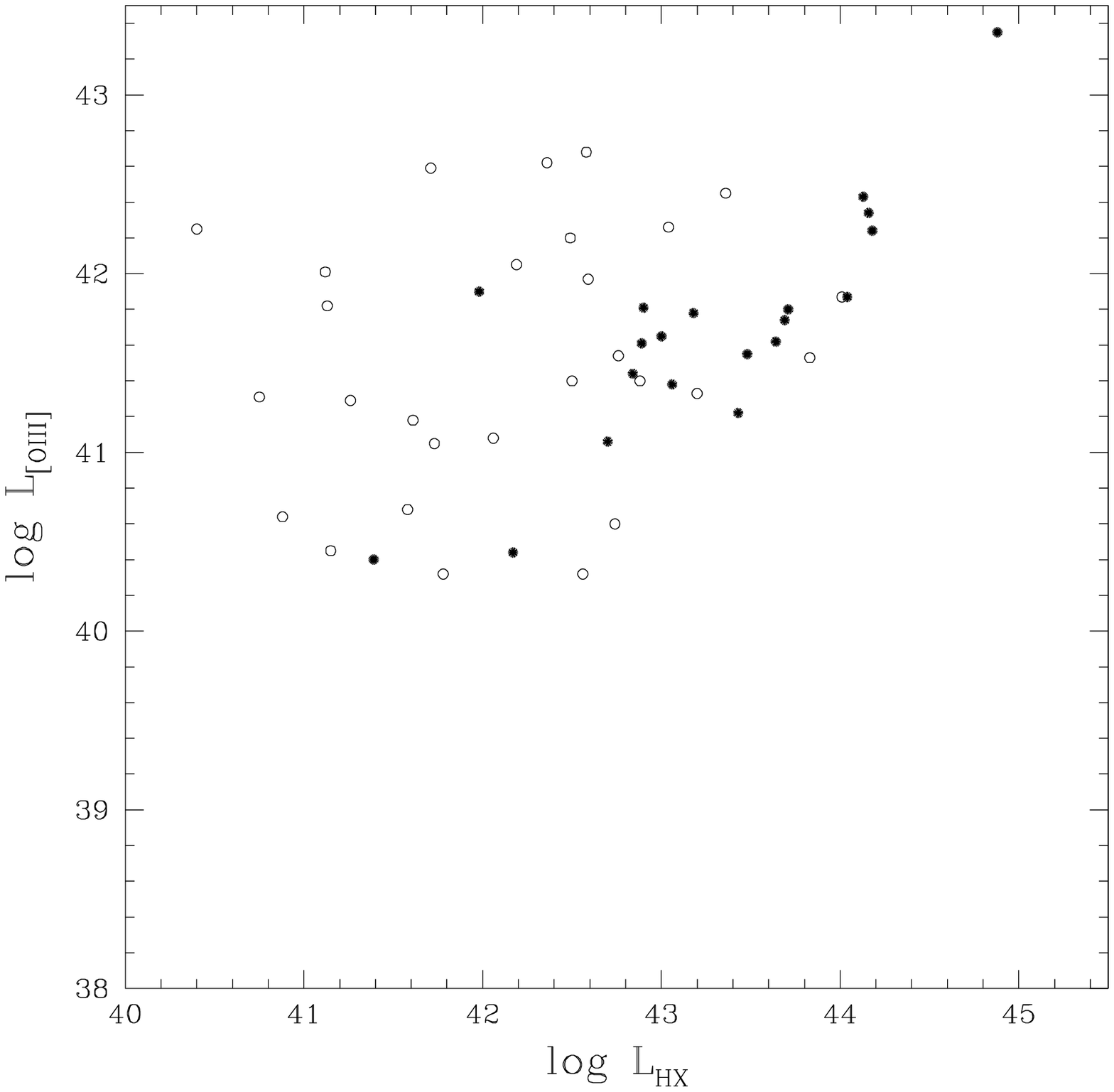}
\caption{A plot of the hard X-ray (2-10 keV) {\it vs.} the [OIII]$\lambda$5007
luminosities for the AGN in Figure 3. The Type 1 AGN are plotted as filled 
circles and the Type 2 AGN as hollow circles. Luminosities are in units of
erg/sec.}
\end{figure}

\begin{figure}
\epsscale{.80}
\plotone{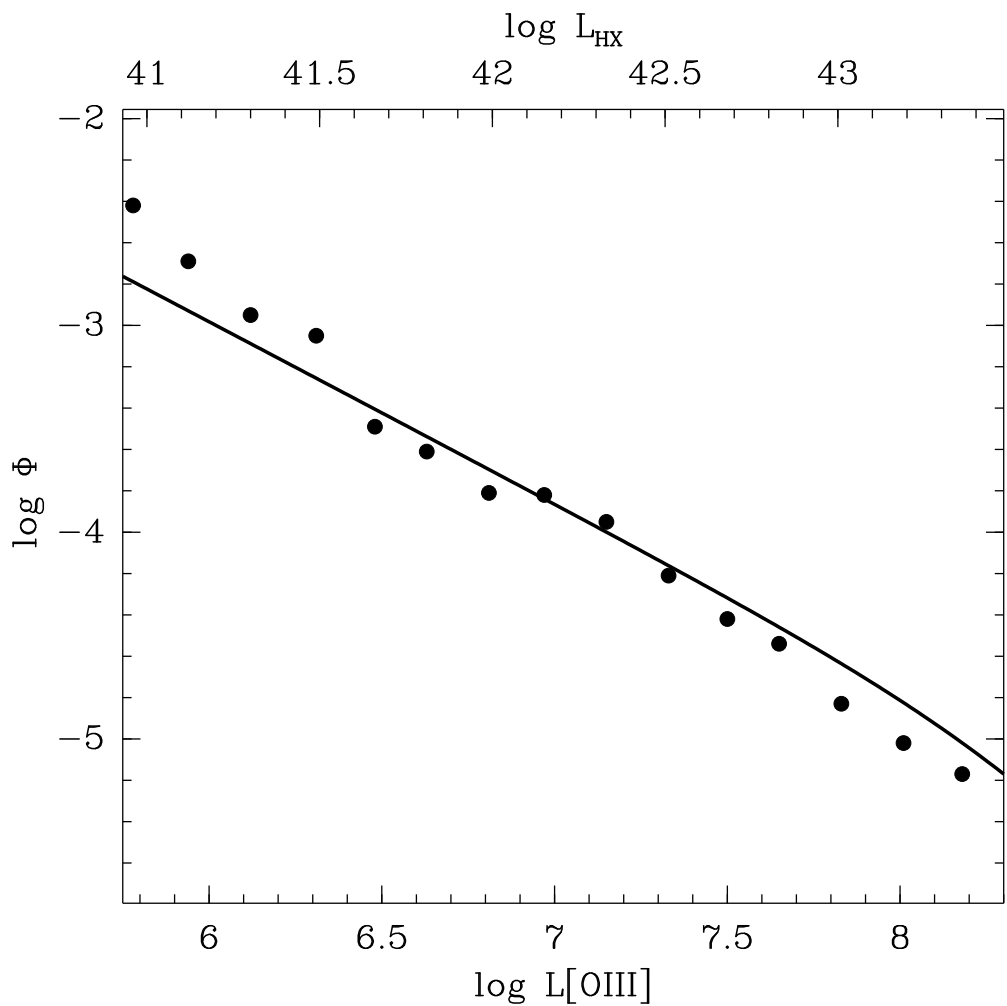}
\caption{A comparison of the local AGN luminosity functions for the 
[OIII]$\lambda$5007 narrow 
emission line (dots - from Hao et al. 2005b) and hard X-ray (3-20 keV)
continuum (line - fit from SR04). Following Hao et al. (2005b) and SR04, the
respective luminosities are in units of $L_{\odot}$ for [OIII] (bottom axis)
and erg/sec for hard X-rays. The two luminosity functions have similar
slopes, and can be aligned in space density for a luminosity
ratio HX/[OIII] = 1.60 dex (as plotted), compared to mean ratio for
hard X-ray selected AGN of 2.15 dex.}
\end{figure}

 
 
 





\clearpage

\begin{deluxetable}{lrrr}
\tablecaption{A Hard X-ray Selected AGN Sample. \label{tbl-1}}
\tablewidth{0pt}
\tablehead{
\colhead{Name} & \colhead{Type}   & \colhead{log$~L_{3-20}$}   &
\colhead{log$~L_{[OIII]}$}}
\startdata
CGCG~$535-012$ & 1 & 43.54 & 41.50\\
NGC~526A & 2 & 43.48 & 41.33\\
Fairall~9 & 1 & 44.14 & 42.05\\
Mrk~590 & 1 & 43.85 & 40.91\\
3C~120 & 1 & 44.22 & 41.87\\
Akn~120 & 1 & 44.06 & 41.34\\
Pic~A & 1 & 43.75 & 41.24\\
NGC~2110 & 2 & 42.88 & 40.35\\
MCG~$+8-11-11$ & 1 & 43.62 & 41.81\\
Mrk~3 & 2 & 43.05 & 42.20\\
Mrk~79 & 1 & 43.51 & 41.65\\
PG~$0804+761$ & 1 & 44.53 & 41.86\\
Mrk~704 & 1 & 43.71 & 41.42\\
Mrk~110 & 1 & 43.86 & 41.82\\
MCG~$-5-23-16$ & 2 & 43.36 & 41.85\\
NGC~3227 & 1 & 42.46 & 40.44\\
NGC~3516 & 1 & 43.10 & 41.06\\
NGC~3783 & 1 & 43.37 & 41.44\\
NGC~3998 & 2 & 41.70 & 39.14\\
NGC~4051 & 1 & 42.10 & 40.40\\
NGC~4151 & 1 & 43.14 & 41.90\\
PG~$1211+143$ & 1 & 44.28 & 42.25\\
NGC~4258 & 2 & 41.10 & 38.71\\
Mrk~205 & 1 & 43.94 & 41.82\\
NGC~4388 & 2 & 42.71 & 40.32\\
3C~273 & 1 & 46.04 & 43.18\\
NGC~4507 & 2 & 42.78 & 41.54\\
NGC~4593 & 1 & 42.63 & 40.53\\
IC~4329A & 1 & 43.97 & 41.22\\
Mrk 279 & 1 & 44.00 & 41.50\\
NGC~5506 & 2 & 43.12 & 40.60\\
NGC~5548 & 1 & 43.71 & 41.62\\
Mrk~841 & 1 & 43.73 & 41.86\\
Mrk~290 & 1 & 43.49 & 41.74\\
NGC~6300 & 2 & 41.68 & 38.86\\
3C~382 & 1 & 44.64 & 41.88\\
3C~390.3 & 1 & 44.45 & 42.43\\
ESO~$140-G043$ & 1 & 42.85 & 41.06\\
Mrk~509 & 1 & 44.26 & 42.34\\
IC~5063 & 2 & 42.83 & 41.40\\
NGC~7314 & 2 & 42.34 & 39.57\\
Mrk~915 & 1 & 43.38 & 41.78\\
Akn~564 & 1 & 43.47 & 41.44\\
NGC~7469 & 1 & 43.33 & 41.55\\
Mrk~926 & 1 & 44.29 & 42.24\\
PG~$2304+042$ & 1 & 43.73 & 41.34\\
NGC~7582 & 2 & 44.01 & 40.32\\
\enddata


\tablecomments{Based on observed hard X-ray fluxes (3-20 keV band) from SR04
and [OIII]$\lambda$5007 data from
X99 and W92 (with no correction for intrinsic dust extinction).
All luminosities are in units of erg sec$^{-1}$, and we take
$H_0 = 70$ km sec$^{-1}$ Mpc$^{-1}$.}

\end{deluxetable}

\clearpage

\begin{deluxetable}{lrrrr}
\tabletypesize{\scriptsize}
\tablecaption{An [OIII]$\lambda$5007 Bright AGN Sample. \label{tbl-2}}
\tablewidth{0pt}
\tablehead{
\colhead{Name} & \colhead{Type}   & \colhead{log$~L_{3-20}$}   & 
\colhead{log$~L_{2-10}$}   & \colhead{log$~L_{[OIII]}$}}
\startdata

NGC~526A & 2 & 43.48 & 43.20 & 41.33\\
NGC~1068 & 2 & $<$41.90 & 41.13 & 41.82\\
NGC~1275 & 2 & $<$43.23 & 43.83 & 41.53\\
NGC~1386 & 2 & $<$42.07 & 39.59 & 40.59\\
NGC~1566 & 1 & $<$42.03 & ----- & 40.11\\
NGC~1685 & 2 & $<$43.10 & 40.40 & 42.25\\
NGC~2273 & 2 & $<$42.52 & 40.88 & 40.64\\
NGC~2992 & 2 & $<$42.52 & 42.06 & 41.08\\
NGC~3081 & 2 & $<$42.58 & 41.26 & 41.29\\
NGC~3227 & 1 & 42.46 & 42.17 & 40.44\\
NGC~3516 & 1 & 43.10 & 42.70 & 41.06\\
NGC~3783 & 1 & 43.37 & 42.84 & 41.44\\
NGC~4051 & 1 & 42.10 & 41.39 & 40.40\\
NGC~4151 & 1 & 43.14 & 41.98 & 41.90\\
NGC~4388 & 2 & 42.71 & 42.56 & 40.32\\
NGC~4507 & 2 & 42.78 & 42.76 & 41.54\\
NGC~5506 & 2 & 43.12 & 42.74 & 40.60\\
NGC~5548 & 1 & 43.71 & 43.64 & 41.62\\
NGC~5643 & 2 & $<$41.95 & 41.15 & 40.45\\
NGC~5728 & 2 & $<$42.73 & 41.61 & 41.18\\
NGC~7212 & 2 & $<$43.59 & 42.19 & 42.05\\
NGC~7469 & 1 & 43.33 & 43.48 & 41.55\\
NGC~7582 & 2 & 44.01 & 41.78 & 40.32\\
Mrk~1 & 2 & $<$43.19 & 40.75 & 41.31\\
Mrk~3 & 2 & 43.05 & 42.49 & 42.20\\
Mrk~6 & 1 & $<$43.80 & 42.90 & 41.81\\
Mrk~34 & 2 & $<$44.15 & 41.71 & 42.59\\
Mrk~78 & 2 & $<$43.88 & ----- & 42.24\\
Mrk~79 & 1 & 43.51 & 43.00 & 41.65\\
Mrk~270 & 2 & $<$42.64 & 41.58 & 40.68\\
Mrk~290 & 1 & 43.49 & 43.69 & 41.74\\
Mrk~348 & 2 & $<$43.09 & 42.88 & 41.40\\
Mrk~463 & 2 & $<$44.15 & 42.36 & 42.62\\
Mrk~477 & 2 & $<$43.89 & 42.58 & 42.68\\
Mrk~509 & 1 & 44.26 & 44.16 & 42.34\\
Mrk~533 & 2 & $<$43.67 & 42.59 & 41.97\\
Mrk~573 & 2 & $<$43.21 & 41.12 & 42.01\\
Mrk~595 & 1 & $<$43.57 & 42.89 & 41.61\\
Mrk~766 & 1 & $<$42.95 & 43.06 & 41.38\\
Mrk~915 & 1 & 43.38 & 43.18 & 41.78\\
Mrk~926 & 1 & 44.29 & 44.18 & 42.24\\
Mrk~1388 & 2 & $<$43.40 & ----- & 41.47\\
IC~4329A & 1 & 43.97 & 43.43 & 41.22\\
IC~5063 & 2 & 42.83 & 42.50 & 41.40\\
Tol~$0109-383$ & 2 & $<$42.88 & 41.73 & 41.05\\
Tol~$1351-375$ & 2 & $<$44.18 & 43.36 & 42.45\\
MCG~$+8-11-11$ & 1 & 43.62 & 43.71 & 41.81\\
Fairall~51 & 1 & $<$43.05 & ----- & 41.06\\
IRAS~$0107-03$ & 2 & $<$44.22 & ----- & 42.44\\
IRAS~$0412-08$ & 1 & $<$43.91 & ----- & 42.02\\
PG~$0119+22$ & 2 & $<$44.19 & 43.04 & 42.26\\
PG~$1612+26$ & 1 & $<$44.98 & 44.88 & 43.35\\
3C~120 & 1 & 44.22 & 44.04 & 41.87\\
3C~390.3 & 1 & 44.45 & 44.13 & 42.43\\
CGCG~$498-038$ & 2 & $<$43.72 & 44.01 & 41.87\\

\enddata

\tablecomments{The hard X-ray data in both the 3-20 keV and 2-10 keV bands
are based
on the observed fluxes (see text and Appendix).
Upper limits in the 3-20 keV band assume the nominal XSS limiting flux of
2.5 $\times 10^{-11}$
erg cm$^-2$ s$^{-1}$ (SR04)..  
The hard X-ray data in the 2-10 keV band come from
the compilations in X99 and Bassani et al. (1999) or from
our analysis of archival data. 
The [OIII]$\lambda$5007 data come from
X99 and W92. They have not been corrected for intrinsic dust extinction.
All luminosities are in units of erg sec$^{-1}$, and we take
$H_0 = 70$ km sec$^{-1}$ Mpc$^{-1}$.}

\end{deluxetable}
 
\end{document}